\documentstyle[12pt]{article}

\def \be{\begin{equation}}
\def \ee{\end{equation}}
\def \bea{\begin{eqnarray}}
\def \eea{\end{eqnarray}}
\begin{document}
{\hskip 9cm IPM - 96}      
\begin{center}
{\Large{\bf{ Logarithmic Correlation Functions in Two Dimensional Turbulence
}}}
\vskip 1 cm   
{{{\bf M.R. Rahimi Tabar}$^{(1,2)}$ , and {\bf S. Rouhani}$^{(1,3)}$ }}
\vskip .5cm
{\it \small {1) Institue for Studies in Theoretical Physics 
and Mathematics
\\ Tehran P.O.Box: 19395-5531, Iran. \\ 2) 
Dept. of Physics , Iran  University of Science and Technology,\\
Narmak, Tehran 16844, Iran.\\3) Department of Physics, 
Sharif University of Technology\\Tehran P.O.Box:11365-9161, Iran  }}
\end{center}
\vskip .1cm
\begin{abstract}
We consider the correlation functions of two-dimensional 
turbulence in  the presence and absence
of a three-dimensional perturbation, by means of conformal 
field theory.
In the persence of three dimensional perturbation, we show that 
in the strong coupling limit of a small scale random 
force, there is
some logarithmic factor in the correlation functions of 
velocity stream functions.
We show that the logarithmic conformal field theory $c_{8,1}$
describes the 2D- turbulence both in the absence and in the presence of the 
perturbation. We obtain the following energy spectrum   
$E(k) \sim k^{-5.125 } \ln(k )$ for perturbed 2D - turbulence and
$E(k) \sim k^{-5 } \ln(k )$ for unperturbed turbulence.
Recent numerical simulation and experimental results confirm our prediction.
\end{abstract} 
\newpage
 \section {{Introduction}} 
{\hskip .5cm}  Polyakov has shown [1] that the exponent of the 
 energy spectrum
of 2D-turbulence can be found by means of conformal field theory. 
 He shows that the energy spectrum behaves as $k^{4\Delta_{\phi}+1}$, 
where $\Delta_{\phi}$
is the dimension of the velocity stream function.
This spectrum is different from the $k^{-3} \ln^{- 1 \over 3}(kL)$
law proposed by Kraichnan [2]. Experimental and numerical simulation results
seem to be even more controversial. Nearly all of the experiments 
have concentrated
on the case of decaying turbulence, which depends strongly on the initial 
conditions [3]. These experiments predict the energy spectrum to
be initially proportional to $k^{-3}$, with the exponent changing with time. 
The case of decaying turbulence has been considered
recently by many authors (see [4] and references therein)
in the context of confomal field theory. 
 Stationary experiments of the two-dimensional turbulence 
 has been considered in [5-7], 
which
show that there is a strong deviation from a $k^{-3}$ spectrum. 
Borue [8] has performed direct numerical
simulation of the 2D-Navier-Stokes equations with a
white noise in time, and 
with non-zero 
correlation in momentum space at some characteristic scale $k_{f}$ 
$( k_{f}\sim {1 \over {L}}$, where L is the scale of system).
The main results of [8] are as follows: both the stirring force and 
dissipation lead to a correction to $k^{-3}$ law, and near $k\sim k_{f}$
the energy spectrum behaves as $E(k)\sim k^{-3} \ln(kL)^{-1\over3}$ which 
is different  from the one-loop result
of Kraichnan [2]. However Falkovich and Lebedov [9] 
have derived Kraichnan's spectrum by using 
Quasi-Lagrangian (QL) variables. 
They predict the correlation of vorticity to be
$<\omega^{n}(r_{1}) \omega^{n}(r_{2})> \sim \ln^{2n \over{3}}
({L \over|r_{1}-r_{2}|}) $, which for $n=1$ gives  the Kraichnan spectrum. 
Here we wish to address the question: how can one find 
an energy spectrum with a logarithmic factor 
in the context of conformal turbulence. 
Recently we have considered the existence of such 
logarithmic factors with integer power, in the energy spectrum of
turbulent 2D - magnetohydrodynamics [10].
In ref.[10] it was shown that, when the Alf`ven effect (i.e. equipartition
of energy between velocity and magnetic modes [11]) is  taken into 
account one is naturally lead to consider, conformal field theories
which have logarithmic terms in their correlation functions.
There has been considerable interest in logarithmic conformal 
field theories (LCFT) [10-21,27]. In these theories there exist atleast two field 
with equal conformal dimensions, such theories admit logarithmic correlation 
functions [12]. 
Recently Moriconi [22] has  considered the problem of conformal 
turbulence taking into account the influence of three-dimensional
effects. He has considered a quasi two-dimensional fluid which is perturbed 
by a small
scale noise representing the effect of the additional degree of freedom
perpendicular to the plane of motion.
Another scheme of perturbation was considered in [23].
Here we consider the perturbed conformal turbulence proposed in [22]
and in the context of  quasi-two-dimensional turbulence, we  first, show  
that there are some constraints similar to the condition imposed by the 
Alf`ven effect. Then we show that these constraints guarantee an 
energy spectrum with a logarithmic factor, and present a solution.\\
The paper is organized as follows; in section 2 we give a brief 
summary of perturbed conformal turbulence and the implication of logarithmic 
terms in correlation functions of velocity stream function.
In section 3 we consider the strong coupling limit of small 
scale external random force perpendicular to plane of motion, and
find the constraints for reducing the number of candidate 
 CFT models. We find a solution within the $c_{p,1}$ series
and derive the energy spectrum and show that the model $c_{8,1}$ 
describe both perturbed and unperturbed 2D-turbulence.
\section  {Quasi Two - Dimentional Turbulence} 
{\hskip .5cm} There are interesting experiments [24], which have 
investigated the 2D 
turbulent fluid, where there was a fluctuating grid responsible for
the perturbation of the two-dinensional motion of the fluid. 
The important observation is that the fluid should be described 
in terms of two-dimensional equations containing not only
the large scale forces but also a small-scale random
perturbation along the direction perpendicular to the direction of motion.
Two-dimensional Navier-Stokes equations take the form,
\be  \partial_t v_\alpha + v_\beta \partial_\beta v_\alpha = \nu \partial^2 
v_\alpha + f_{\alpha} ^{(1)} + g f_{\alpha} ^{(2)} - \partial_\alpha P
\ee
where $v$ is the velocity field, and $\nu$ is the viscosity and 
 $f_{\alpha} ^{(1)}(x,t)$ and $f_{\alpha } ^{(2)}(x,t)$ are 
 the stirring forces 
 defined at large scales
L and medium scales $a<<y<<L$, respectively, $a$ is the dissipation scale.
An appropriate correlation for $f_{\alpha } ^{(1)}$ is
\be
<f_{\alpha } ^{(1)} ({\bf{x}},t) f_{\alpha } ^{(2)}({\bf{x^{'}}},t^{'})> = 
\delta_{\alpha \beta} k(0)(1- \delta
(({\bf{x-x^{'}}})^2-L^2)) \delta (t-t^{'}) 
\ee
where $\alpha, \beta =1,2$ , $f_3 ^{(1)}=0$ and $f_1 ^{(2)}=f_2 ^{(2)}=0, 
f_3 ^{(2)} \neq 0$.
The dimensionless constant $g$ shows a coupling with the three-dimensional 
modes of the fluid. As pointed out in ref.[20,24], 
when we have some external noise along the direction perpendicular to the 
direction of motion one has
to be careful of the compressibility condition for velocity field in two
dimensions. When  considering the  three-dimensional velocity
field, compressibility condition is $\partial_{i}v_{i}=0$, 
where $i=1,2,3$. If we project this constriant to the two-dimensional 
plane it follows that $\partial_\alpha v_\alpha=O(g)$, 
$\alpha=1,2$. Therefore to take this point into account, the 
velocity field may be written as 
\be 
v_\alpha = \epsilon_{\beta \alpha} \partial_\beta \psi + g \partial_
\alpha \phi 
\ee
Where $\psi$ and $\phi$ are the velocity stream function and the velocity 
potential, respectively. The divergence of two dimentional velocity 
field is $\rho = g\partial^2\phi$. 
Following [22],  we expand $\psi$ and $\phi$ in powers of $g$ in 
the following forms:
\bea  
\psi=\sum^\infty_{n=0} g^n \psi^{(n)}  \hskip 0.5cm \omega=\sum^\infty
_{n=0} g^n \omega^{(n)} \\  
\phi= \sum^\infty_{n=0} g^n \phi^{(n)}
\hskip 0.5cm \rho=\sum^\infty_{n=0} g^{n+1} \rho^{(n)} 
\eea
By taking curl of both sides of eq.(1) and substituting eqs.(4) and (5),
one can find exact infinite chain of equations for $\omega^{(n)}$
and $\rho^{(n)}$ as follows.
\bea
\partial_t \omega^{(n)} &+ \sum^n_{p=0} \epsilon_{\alpha
\beta}
\partial_\alpha \psi^{(p)}\partial_\beta \partial^2 \psi^{(n-p)} +
\sum^{n-1}_{p=0} \left[ \partial_\beta \phi^{(p)} \partial_\beta
\partial^2 \psi^{(n-p-1)} +
\partial^2 \phi^{(p)} \partial^2 \psi^{(n-p-1)} \right] \nonumber \\
&= \nu \partial^2 \omega^{(n)} + \epsilon_{\alpha \beta} \partial_{\alpha}
f^{(2)}_\beta \delta_{n,1} \ , \
\eea
\bea 
\partial_t \omega^{(0)} &&+ \epsilon_{\alpha \beta}
\partial_\alpha \psi^{(0)}\partial_\beta \partial^2 \psi^{(0)} =
\nu \partial^2 \omega^{(0)}+\epsilon_{\alpha \beta} \partial_{\alpha}
f^{(1)}_\beta \ 
\eea
\bea
&\partial_t \rho^{(n)} + \sum^{n-1}_{p=0}
\left[ \partial_\alpha \partial_\beta \phi^{(p)}
\partial_\alpha \partial_\beta \phi^{(n-p-1)} +
\partial_\alpha \phi^{(p)} \partial_\alpha \partial^2 \phi^{(n-p-1)}
\right] 
 \nonumber\\
&+\sum^n_{p=0} \left[2 \epsilon_{\alpha \beta}
\partial_\beta \partial_\sigma \phi^{(p)}
\partial_\alpha \partial_\sigma \psi^{(n-p)}
+\epsilon_{\alpha \beta} \partial_\alpha \psi^{(n-p)} \partial_\beta
\partial^2\phi^{(p)}  \right] 
 \nonumber\\
&+\sum^{n+1}_{p=0} \left[ \partial_\alpha \partial_\beta \psi^{(p)}
\partial_\alpha \partial_\beta \psi^{(n-p+1)}
- \partial^2 \psi^{(p)} \partial^2 \psi^{(n-p+1)} \right]
= \nu \partial^2 \rho^{(n)} & 
\eea
\bea 
&\partial_t \rho^{(0)} +2\partial_\alpha \partial_\beta
\psi^{(0)}\partial_\alpha \partial_\beta \psi^{(1)}
+2\epsilon_{\alpha \beta}\partial_\beta \partial_\sigma \phi^{(0)}
\partial_\alpha \partial_\sigma \psi^{(0)}+
\epsilon_{\alpha \beta} \partial_\alpha \psi^{(0)} \partial_\beta
\partial^2\phi^{(0)} 
  \nonumber\\
&-2 \partial^2 \psi^{(0)} \partial^2 \psi^{(1)}
=\nu \partial^2 \rho^{(0)}+\partial_\alpha f^{(2)}_\alpha 
\eea
Eq.(7) is identical to the case of an unperturbed 
(i.e. $g=0$) two-dimensional turbulent fluid.
This means that the field $\psi^{(0)}$ will be related to an 
enstrophy or energy cascade, even in the presence of three dimensional
fluctuations. In other words the enstrophy and energy 
cascade conditions do not change in the presence of
3D- perturbation. We will consider this important point in the end of this 
next 
section.
The basic assumption of Moriconi [22], is that not only $\psi^{(0)}$ 
but also
the other components in the power expansions of $\psi$ and $\phi$ 
are primary 
operators which belong to some conformal field theory.
There is an important point here: noting that $g$ is a dimensionless 
coupling constant eqs. (4) and (5) tell us all the  
$\psi^{(n)}$`s and $\phi^{(n)}$`s have the same scaling dimension.
To avoid this difficulty it has been suggested [22] that 
there is a hidden scale $l$ in the problem which may be related to 
the intermittency effect. Therefore expansions in the eqs. (4) and (5) 
change to 
\bea
\psi=\Sigma f_{n} l^{2\Delta_{\psi^{(n)}}} g^n \psi^{(n)} 
\\  \nonumber
\phi=\Sigma f_{n}^{'} l^{2\Delta_{\phi^{(n)}}} g^n \phi^{(n)}
\eea
where $l$ is some scale proportional to  
$\sim \nu^\alpha <\omega^2>^\beta <v^2>^\gamma$
where $2\alpha + 2\gamma = 1$ and $\alpha + 2\beta +2\gamma = 0$.
If one accepts this prescription still there is  some 
ambiguity in the determination of the energy spectrum exponents, as one can 
select all of the exponents 
$4\Delta_{\phi^{(n)}} + 1$ and $ 4\Delta_{\psi^{(n)}} + 1$ as the exponents 
of the energy 
spectrum. In this situation we have to use a CFT with an infinite number of
primary fields. However if we restrict ourselves to the strong coupling 
limit, a finite number of primary fields suffices as can be seen from 
eq.(13) below. 

Furthermore there is the possiblity of some fields $\psi^{(n)}$ and
$\phi^{(n)}$ in the expansions (10) having the same scaling dimension 
this leads to logarithmic correlators.
According to Gurarie [12] if the operator product expansion $(OPE)$ of some 
fields in $CFT$ model possess at least two operators with the same 
dimension, one naturally gets logarithmic correlation functions. 
 In other words the operators with the 
same scaling dimension form the basis of the Jordan cell for  $L_{0} [12]$.
Here we assume that two or more operators form the basis of the Jordan
cell for operator $L_0$ i.e. they have equal dimensions, then: 
\bea
 L_{0} \Psi^{(n)}&=& \Delta_{\Psi^{(n)}} \Psi ^{(n)} +\Psi^{(n-1)}  \hskip 1cm n>0  
\nonumber \\
 L_{0}\Psi^{(0)}&=&\Delta_{\Psi^{(0)}} \Psi^{(0)} 
\eea
where $\Psi^{(n)}$ may be any of $\psi^{(n)}$ or $\phi^{(n)}$. Therefore 
the standard 
$OPE$ [12] takes the following form:
\be
 \Phi^{(n_{1})}(z) \Phi^{(n_{2})}(0) = z^{2(\Delta_{\Psi^{(m-n)}} - \Delta_
{\Psi^{(n_2)}} - \Delta_{\Psi^{(n_1)}})} 
(\Sigma_n \ln^{n}(z) {\Psi^{(m-n)} + decendents })
\ee
This argument shows that, we have to consider logarithmic conformal
field theories as candidates for describing such systems.
To find the explicit form of the energy spectrum, we concentrate on 
the strong 
coupling limit.
\section {Strong Coupling Limit and The Logarithmic Correlation}
{\hskip .5cm} As discussed in [22], by only considering $\psi^{(0)}$ 
and $\psi^{(1)}$
and $\phi^{(0)}$, we can completely describe the strong coupling limit of 
eqs.(1). Noting that if the  
constant flux condition is not satisfied by the pair of fields 
$\psi^{(0)}$ and 
$\phi^{(0)}$, then there exists no further soultions for the 
model under consideration. 
Therefore it is enough to consider 
those  models for which the field $\psi^{(0)}$ 
satisfies the nonperturbative constraint and  
$\psi^{(1)}$ and $\phi^{(0)}$ satisfy the constraints associated 
with the three-dimensional effect. Taking this into account  
, the expansions of $\psi$ and $\phi$ can be written as 
\bea
\psi&=&\psi^{(0)} + f_{a}(g) \psi^{(1)} 
\nonumber \\
\phi&=&f_{b}(g) \phi^{(0)} 
\eea
where $f_{a}(0)=f_{b}(0)=0$.
Following [22] we consider three limits for $f_{a,b}$
\bea  
a)\hskip .3cm &  g\rightarrow 0 \hskip 1cm & f_{a,b} \rightarrow 0  
\nonumber \\ 
b)\hskip .3cm &  g\rightarrow 0 \hskip 1cm & f_{a,b} \sim 1  
\nonumber \\ 
c) \hskip .3cm & g\rightarrow \infty \hskip 1cm & f_{a,b} \rightarrow diverges 
\eea
the case (c) may be defined as the strong coupling regime.
It is clear that in the case of the strong coupling, 
the contribution of $\psi^{(0)}$ field can be discarded.
However in the presence of any perturbation 
the constant enstrophy cascade condition depend only on $\psi^{(0)}$,
which as mentioned follows fron eq.(7).
It has been shown in [22], that eq.(14-a) dose not give any 
correction to the power law spectrum. 
Careful consideration shows that, the other two cases lead to a logarithmic 
factor in the energy spectrum. In the strong coupling limit, 
where the inertial range exponent derived 
from $\psi^{(0)}$ may be discarded and the exponent 
can be determined by considering the the dimension of $\psi^{(1)}$ and 
$\phi^{(0)}$.
Here we can apply the same argument as that of [11]. 
Applying the dynamical renormalization group one can show
that by the existence of a critical dynamical index
of eqs. (6), (7) and (9) for $\psi^{(1)} $ and $\psi^{(0)}$
and $\phi^{(0)}$, leads to condition that 
\be
\Delta_{\psi^{(0)}}=\Delta_{\psi^{(1)}}=\Delta_{\phi^{(0)}}
\ee
This means that in the steady state there is some 
equipartition 
of energy between different components of the velocity field [11].

However in the strong copling limit 
the contribution of $\psi^{(0)}$
field can be discarded therefore we have 
 two fields (i.e $\psi^{(1)}$ and $\phi^{(0)}$) in some CFT 
model and one naturally
obtains a logarithmic factor 
in the energy
spectrum. 
 The main idea is as follows.\\
the operator product expansion of two fields $A$ and $B$,
which have two fields
$\phi^{(0)}$ and $\psi^{(1)}$ of equal dimensions in their fusion rule, has a logarithmic
term:
\be
A(z) B(0) = z^{h_\phi - h_A - h_B}\{\psi^{(1)} (0) + \ldots +\log z(\phi^{(0)} (0) + \ldots )
\}
\ee
to see this it is sufficient to look at the four-point function :
\be
<A(z_1) B(z_2) A(z_3) B(z_4)> \sim {1\over {(z_1-z_3)}^{h_A}} {1 \over
{(z_2 - z_4)}^{h_B}} {1 \over {[x(1-x)]^{h_A + h_B - h_\phi}}}
F(x)
\ee
Where the cross ratio $x$ is given by :
\be
x = {{(z_1 - z_2)(z_3 - z_4)}\over{(z_1 - z_3)(z_2 - z_4)}}
\ee
In degenerate minimal
models $F(x)$ satisfies a second order linear differential equation.
Therefore a solution for $F(x)$ can be found in terms of a series expansion :
\be
F(x) = x^\alpha \sum a_n x^n
\ee
It can be easily shown that the existence of two fields with equal dimensions
in OPE of $A$ and $B$ is equivalent to the secular equation for $\alpha$
having coincident roots [9], in which case two
independent solutions can be constructed according to :
\be
\sum b_n x^n + \log x \sum a_n x^n
\ee
Now consistency of equations (16) and (20) requires :
\be
<A(z_1) B(z_2) \psi(z_3)> = <A(z_1) B(z_2) \phi(z_3)>\{ \log{(z_1-z_2) \over
{(z_1-z_3)(z_2-z_3)}} + \lambda \}
\ee
\be
<\psi(z) \psi(0)> \sim {1 \over {z^{2h_\psi}}} [\log z +\lambda^{'}]
\ee
\be
<\psi(z) \phi(0)> \sim {1 \over {z^{2 h_\phi}}}
\ee
where $\lambda$ and $\lambda^{'}$ are constants.
The IR-problem of such system has been discussed in [9].
The energy spectrum of this type of correlations has following:
\be
 E(k)\sim k^{-4 |\Delta_{\phi^{(0)}}|+1} \ln (kL)
\ee
Let us rewrite the constant enstrophy condition, in order to find
 a logarithmic 
CFT for 2D - turbulence.
Consider the fusion of field $\psi^{(0)}$ with itself:
\be
\psi^{(0)} \times \psi^{(0)} = \chi+\cdots
\ee
such that $\chi$ is the field with minimum conformal dimension , on the right 
hand side.
Then the constant enstrophy condition implies:
\be
\Delta_{\psi^{(0)}} + \Delta_{\chi}=-3 \hskip 1cm and \hskip 1cm 
\Delta_{\chi} > 2 \Delta_{\psi^{(0)}}
\ee
According to [22] field $\psi^{(0)}$ will be related to an enstrophy 
cascade, even in the presence of three - dimensional effect.
A possible candidate may exist within the $c_{p,1}$ series [19,25].
The central charge for this series is $c=13-6(p+p^{-1})$.
This series is special since it has $c_{eff}=1$.
These CFT`s posess $3p-1$ highest weight representation with conformal
dimensions:
\be
h_{ps}=\frac{(p-s)^2-(p-1)^2}{4p} \hskip 1cm  1 \leq s \leq 3p-1
\ee
of these $2(p-1)$ have pair wise equal dimensions.
Two field $\phi_s$ and $\phi_{s'}$ have equal and negative weights provided 
that $s+s'=2p,(s \neq 1,2p-1)$.
Let us adopt such a pair as candidates for  $\psi^{(1)}$ and  $\phi^{(0)}$.
We can now look for the candidate values of $s$ such that eq.(26) is satisfied.
The only solution is given by $p=8$ with $c= - \frac{286}{8}$,
where the set of fields with negative dimensions are:
\be
( -\frac {3}{2}, -\frac {3}{2}, -\frac {3}{4}, -\frac {3}{4}, -\frac {5}{4},
-\frac {5}{4}, -\frac {13}{32}, -\frac {13}{32}, -\frac {33}{32},
-\frac {33}{32}, -\frac {45}{32}, -\frac {45}{32}, -\frac {49}{32},-\frac {49}{32})
\ee
With $\Delta_{\psi^{(0)}}=-\frac{3}{2}$. 
Therefore for unperturbed turbulence we have:
\be
E(k) \sim k^{-5 } \ln(k )
\ee
For perturbed turbulence  $\phi^{(0)}$ and $\psi^{(1)}$ can be assigned as 
any pair from this LCFT. Different choices lead to different exponents for the 
energy spectrum, these are:
\be
 (-2, -4, -0.625, -4.025, -5.125, -3.125)
\ee
therefore the best exponent to fit the experimental data [22,7] is $-5.125$ 
which corresponds to conformal weights:
\be
\Delta_{\phi_{(0)}} = \Delta_{\psi^{(1)}}=-\frac{49}{32}
\ee
The energy spectrum thus is given by:
\be
E(k) \sim k^{-5.125 } \ln(k )
\ee
which confirms witn the numerical analysis by Borue [8], and 
experimental date [22,7].  \\
Furthermore we can relax the conditions and ask
     which types of conformal 
    field theory may be used for
modeling of 2D - turbulence , provided we assume the condition 
\be
\Delta_{\phi^{(0)}} \simeq \Delta_{\psi^{(1)}}
\ee
as well as the cascades of 
constant 
enstrophy and the constant energy. 
In [9], it has been shown that the standard minimal models,
do not have logarithmic correlators 
becuase for  (p , q) coprime, all the primary fields in minimal 
models have different dimensions. 
However almost logarithmic behavior is 
obtained when two primary fields have almost equal conformal dimensions.
Where we have $|\Delta \phi^{(0)} - \Delta \psi^{(1)}| = \epsilon$, 
and $\epsilon$ satisfies [10]:  
\be
\epsilon \leq \frac{1}{5/2 \log R_{e}}
\ee
provided $z$ lies in inertial range:
\be
a << z << R
\ee
where $R_{e}$ is the Reynold's number of system,
$a$ and $R$ are the dissipation and the large scales of
the system repectively.
The table of CFT models are nearly consistent with eq.(25)
is given in reference [22]. 
Eq.(34) is the relation of the dimensions of fields $\psi^{(1)}$ and $\phi^{(0)}$ 
and the Reynold`s number of system. \\
{\bf Acknowledgements:} We would like to thank Mehran Kardar for 
valuable discussions about investigation of the Navier-Stokes equations 
with some noise and Werner Nahm for comments on the  
origin of logarithmic factors in the correlation functions.

\end{document}